\documentclass[12pt]{iopart}
\usepackage{amssymb}
\usepackage{graphicx}
\DeclareGraphicsExtensions{.eps,.ps,.eps.gz,.ps.gz,.eps.Z}
\usepackage{dcolumn}
\usepackage{bm}

\begin{document}

\title[Local capacitance of a QH bar]{Classical and quantum capacitances calculated locally considering a two-dimensional Hall bar}
\author {E. Guvenilir$^{a} $, O. Kilicoglu$^{b} $, D. Eksi$^{c}$ A. Siddiki$^{d,*}$ }
\address {$^{a} $Faculty of Sciences and Letters, Department of Physics, Istanbul Technical University, Istanbul 34460, Turkey}
\address {$^{b} $Faculty of Sciences and Letters, Department of Physics, Isik University, Istanbul 34980, Turkey}
\address {$^{c} $Vocational School of Health Services, Yeni Yuzyil University, 34010 Istanbul, Turkey}
\address {$^{d} $Faculty of Sciences and Letters, Department of Physics, Mimar Sinan Fine Arts University, 34380-Sisli, Istanbul, Turkey}
\ead{afifsiddiki@gmail.com}
\vspace{10pt}
\today

\begin{abstract}
In this work we investigate the electrostatic properties of two dimensional electron system (2DES) in the integer quantum Hall regime. The alternating screening properties of compressible and incompressible strips are formed due to edge effects together with electron-electron interactions. As it is well known, the Landau quantization emanates from strong perpendicular magnetic fields. The (Landau) energy levels are broadened due to impurities, which we embedded their effects in density of states (DOS). In a basic level DOS has two different forms: the Gaussian and semi-elliptic descriptions. The second form is calculated within the self consistent Born approximation (SCBA). Having in hand the density of states, we obtain both the longitudinal and Hall (transversal) conductivities ($\sigma_{l}, \sigma_{H}$) utilizing Thomas-Fermi-Poisson approximation to calculate position dependent charge density profile and use Drude formalism to obtain transport coefficients. Since, the definition of capacitance is closely related with compressibility via DOS, (local) screening properties of 2DES is extremely important to understand local capacitances. Here we numerically simulate a translational invariant Hall bar subject to high magnetic fields which is perpendicular to the plane of the 2DES using realistic parameters extracted from the related experiments. Using the above mentioned approaches the local capacitances are calculated, numerically. Our findings are in perfect agreement with related experimental results which are based on a dynamic scanning capacitance microscopy technique.
\end{abstract}

\section{Introduction}
Since the first experimental observation of the integer quantized Hall (QH) effect~\cite{vK80:494}, theories explaining the experimental results divide into bulk ~\cite{Laughlin81,Niu87:2188,Hasan10:3045} and edge models~\cite{Buettiker88:9375,Chang90:871,Chklovskii92:4026,Lier94:7757,Oh97:108}  which are discussed within different theoretical frameworks. Starting from late 80's, experimental investigations using capacitive methods shed light on the bulk compressibility/incompressibility of the 2DES~\cite{Weiss90:221,Suddards12:083015}. One of the exciting effects of the magnetic field $B$ on the 2DES is to separate the system into two regions with opposite capacitive and compressibility properties. To be explicit, the compressibility is described as, $\Gamma=n^{-2}_{\rm el}D_{T}$ and quantum capacitance (per area $A$) is given by, $e^2D_{T}$, where $D_{T}$ being the thermodynamic density of states (TDOS), $n_{\rm el}$ is the electron number density per $A$ of a homogeneous and infinite 2DES. TDOS includes both the (geometrical and/or magnetic field induced) confinement dependent DOS, the quantum statistics (Fermi or Bose) of the particles and thermodynamic quantities of the system (e.g chemical potential, temperature etc.). The total capacitance ${C_{\rm T}}$ is obtained by serially adding both classical (geometrical) and quantum capacitances, namely;
\begin{equation}
	\frac{1}{C_{\rm T}}=\frac{1}{C_{\rm geo}}+\frac{1}{C_{\rm q}}\label{Eq:1}
\end{equation}
The most commonly known is the compressible state considered as metallic-like, where DOS is highly degenerate. Recall that in a 3D metal all, screening and transport properties are defined by the Fermi surface and the DOS diverges to infinity. Even if an external $B$ field is applied, DOS is not quantized. However, for a 2D electron system (2DES) DOS is finite due to dimensional constrictions and in the absence of an external $B$ field, DOS is independent of energy and is constant $D_0=m^*/(\pi \hbar^2)$, which only depends on sample parameters (i.e effective electron mass $m^*=0.067m_{\rm el}$, for GaAs, $m_{\rm el}$ being the bare electron mass). As a direct consequence, it is always possible to find a finite number of state at the Fermi energy ($E_{F}\propto n_{\rm el}$). In contrast, the presence of an external magnetic field DOS is Landau quantized (see e.g. Ref. ~\cite{Girvin00:book}) and depends on energy $E$ which we denote as $D_{\rm B}(E)$. Therefore, it is possible to have situations where $E_{F}$ is not equal to a particular Landau energy $E_n$, where $n$ is the index of Landau level and is described as, $E_n=\hbar\omega_c(n+1/2)$, $\omega_c=eB/m^*c$ being the cyclotron frequency. If $E_{F}\neq E_n$, as a consequence $D_{B}(E_{\rm F}=0)$, the system is called incompressible. In the sense that it is not possible to add an extra particle to the system (2DES) without paying relatively high energy. At vanishing temperature and with out any impurities, the required energy approximates to infinity (similar to a Topological insulator~\cite{Hasan10:3045}).
\newline Considering a finite 2DES that is bounded by physical edges in $x$ direction and has translational invariance in $y$, $ n_{\rm el}$ depends locally on the total electrostatic potential $V(x)$ (energy)~\cite{Chklovskii92:4026,Chklovskii93:12605,Fogler94:1656}. The position dependent electron number density $ n_{\rm el}(x)$ can either be obtained by self-consisted calculations~\cite{Wulf88:4218,Guven03:115327,Siddiki04:condmat,siddiki2004}  or by analytical expressions~\cite{Chklovskii92:4026,Siddiki10:67010,Salman13:203}. Spatially varying total potential is determined essentially by the boundary conditions and charge (electron and donor) distributions via Poisson equation. Imposing specific boundary conditions, it is possible to obtain the solution of Poisson equation analytically ~\cite{Chklovskii92:4026,Lier94:7757,Siddiki03:125315}. However, usually assumptions made regarding the sample geometry does not always coincide with real experimental structures. 

On the experimental side, local transparency~\cite{Weitz00:247,Ahlswede01:562,Ahlswede02:165,Dahlem10:121305,Panos:Diss}, compressibility~\cite{Weitz00:247}, transport~\cite{huels04:085319}  and capacitance~\cite{Suddards12:083015} were investigated almost for more then two decades. All the mentioned experimental results converge to the same conclusion that, there exists compressible and incompressible regions, and their widths, spatial location etc. depend on magnetic field strength and physical boundary conditions. The early Single Electron Transistor~\cite{Wei98:1674} (SET) and Scanning Force Microscope (SFM)~\cite{Ahlswede01:562,Ahlswede02:165,Dahlem10:121305,Panos:Diss} experiments clearly show that, due to bending of the Landau levels at the edges stemming from physical boundaries, the incompressible regions are surrounded by compressible ones, if the system is in a quantized Hall state. As one of the main experimentally controllable parameter is $B$ field, in turn, determining whether the system is in a QH state or not. In SET experiments~\cite{Wei98:1674} spatial distribution of edge (incompressible) strips are investigated both varying the $B$ field and the voltage applied to the edge electrode (metallic gate), as another sensitively controllable parameter. It is observed that the locations of incompressible strips are strongly influenced by adjusting these parameters. In SFM experiments, considering narrow Hall bars ($W\lesssim$ 20 $\mu$m)~\cite{Ahlswede02:165,Dahlem10:121305} and only varying the $B$ field it is reported that, the transparency (screening ability of the electrons) strongly depends on the position of incompressible strips while within these regions the external potential is poorly screened and as a result regions become transparent to total potential. These experiments~\cite{Ahlswede02:165,Dahlem10:121305}  are performed on GaAs/AlGaAs heterostructures have higher mobility samples compared to that of SET ones. The results were consistent with the previous ones, however, due to improved experimental setup F. Dahlem et.al could also resolve the hot spots near contacts~\cite{Dahlem10:121305}. Most recently, K. Panos using SFM the technique, measured similar features on Graphene samples~\cite{Panos:Diss}.  

In another method, developed by Shayegan's group utilizes Microwave  impedance microscopy (MIM) to probe the local conductivities of the bulk and edge states at low-temperatures ($T\sim 2$ K)  focusing on filling factor $\nu \simeq2$ ~\cite{Lai11:176809}, using GaAs/AlGaAs heterostructures similar to Stuttgart group. As a brief definition, filling factor is determined by the number of Landau levels below the Fermi energy. If $\nu$ is an integer $k$, it means that the Landau level is fully occupied and the state is incompressible. Otherwise, $\nu\neq k$, the top most $k^{\rm th}$ level is partially occupied, compressible. Whereas all other levels below are completely full. It is important to note that, since the electron density varies depending on location, it is also possible to define a local filling factor $\nu (x)$. Based on the observations performing MIM experiments, K. Lai et. al., reports that their results are quantitatively coinciding with the non-self consistent, zero temperature CGS~\cite{Chklovskii92:4026} model. In deed, the comparison between experimental and electrostatic model can not be compared directly, while the CGS model does not describe realistically the following quantities: \textit{(i)} the DOS is assumed to be infinite within the compressible strips, which is unrealistic for a 2DES where DOS is finite, \textit{(ii)} the analytical expression that describes $n_{\rm el}(x)$ considers a heterostructure where electrons, donors and metallic side gates reside on the same $xy$ plane and \textit{(iii)} transport coefficients are not discussed at all, besides a hand waving statement that compressible regions are metal-like. Therefore, CGS model claims that the current flows along the compressible regions, however, no analytical expressions are given. Hence, making such a quantitative comparison underestimates the properties of local conductivities. In an improved non-self-consistent electrostatic approach by Fogler and Shklovskii~\cite{Fogler94:1656} it is shown that the external (i.e. imposed non-equilibrium) current flows along the incompressible strips where local electrochemical potential $\mu (x)$ presents variations across these strips, thus a drift current is confined within the incompressible strips. Meanwhile, the surrounding compressible regions, due to their relatively good screening property (note that, these regions are not metal with perfect screening), flattens $\mu(x)$ across these regions.  We will discuss in detail the screening properties of the 2DES considering within an improved self-consistent model~\cite{Guven03:115327,siddiki2004}, which also takes into account realistic collision broadened DOS, sample properties (e.g.  mobility and edge effects) at finite temperatures. The latter approach also includes the finiteness of Landau wave-functions, which covers a more general situation beyond Thomas-Fermi Approximation (TFA).

In this paper we focus on an experimental setup that utilizes capacitive coupling of the probing AFM tip and the sample. This technique is developed at Nottingham University and, is named as dynamical scanning capacitance microscopy (DSCM)~\cite{Baumgartner09:013704}. This setup enables measuring the local impedance on quantum Hall bars~\cite{Suddards12:083015}. The electronic setup essentially is composed of two Lock-in Amplifiers measuring both the real (resistive) and imaginary parts (capacitive) of the impedance. Equipped with such information they were able to extract local capacitances. Following the pioneering works of Gerhardts ~\cite{Lier94:7757,Oh97:13519,Guven03:115327,siddiki2004}, we obtained both local electron densities using finite DOS. Granted with these, it is shown that local capacitances depends whether 2DES is compressible or incompressible, locally. Our results are in perfect agreement with the experimental findings. 

This work is organized as follows. First we discuss the electrostatic approximation within the Thomas-Fermi-Poisson approach (TFPA) in Sec.~\ref{Sec.2}. Then, we discuss two different forms of DOS considering a 2DES subject to high magnetic fields, perpendicularly. Sec.~\ref{Sec.3} briefly discusses the experimental techniques and efforts related with the measured impedances. Next the quantum and classical capacitances are discussed within the local equilibrium model. We close our paper by a Conclusion Section, including open questions and also proposing further experimental investigations to test our predictions.

\section{Model}\label{Sec.2}
\subsection{Semi-classical Electrostatic Model}
 As a starting point we first investigate electrostatic properties of a 2DES also taking into account bare Coulomb interaction. Screening properties are already discussed by Wulf et.al.~\cite{Wulf88:4218} numerically, by Chang~\cite{Chang90:871} phenomenologically, and within the non-self-consistent CGS model~\cite{Chklovskii92:4026}, analytically. However, non of these investigations provide a self-standing model to describe QHE. In CGS approach, the 2DES, metallic gates and the positive background charges are assumed to reside all on the same plane, $ z=0 $, which only considers one half of the sample. In our approach, instead, the distance from metallic gates to the 2DES is obtained via Gerhardts's model~\cite{Lier94:7757,Oh97:13519}. This approach neglects density fluctuations due to impurities and lead a monotonically varying 2DES. Later, this model is extended to both long-range potential fluctuations~\cite{Siddiki:ijmp} and short range impurity scattering situations~\cite{Siddiki:Diss}. However, the results obtained by assuming monotonic electron and homogeneous donor distributions, does not differ from the results including fluctuations and impurity scattering, qualitatively.  Due to charge neutrality the average number density of electrons are equal to the positive background charges $n_0$. The boundaries of 2DES, in the interval $-d<x<d$ are defined by applying a negative voltage $-V_g $ to the metallic gates, where $d$ is the half width of the sample. Heterostructure below the 2DES $ (z<0) $ is assumed to be a semiconductor with a dielectric constant $ \kappa \gg 1$ (for GaAs $\kappa\simeq12.4$). Hence, the electrostatic problem can be treated as one dimensional if translation invariance is assumed in $y$ direction. Solution of Poisson equation, with the above given boundary conditions yields an effective total potential (energy) $V(x)$ that electrons experience,
\begin{equation}
V(x)=V_{\rm bg}(x)+V_{\rm H}(x)+V_g ,
\end{equation}
where, $ V_{\rm bg}(x) $ is background potential generated by donors and $V_{\rm H}(x) $ is Hartree potential emanating from bare Coulomb (charge-charge, without spin) interaction and is,
\begin{equation}
V_{\rm H}(x)=\frac{2e^2}{\kappa}\int_{-d}^d dx^{'} K(x,x^{'}) n_{\rm el}(x^{'}),
\end{equation}
$K(x,x')$ is kernel function that solves Poisson equation calculated analytically and is given by,
\begin{equation}
K(x,x^{'})= \ln\left| \frac{\sqrt{(d^2-x^2)(d^2-x^{'2})}+d^2-x^{'}x}{(x-x^{'})d}\right|.
\end{equation}.
It is important to re-emphasize that, due to translational invariance in $y$ direction, reduces potential, density and kernel functions to one dimension, e.g $V(\vec{r})=V(x).V_y$, with $V_y$ being a constant along the sample.

Assuming homogeneously distributed donor density and imposing above described boundary conditions determines the background potential as,
\begin{equation}
V_{\rm bg}(x)=-E^0_{\rm bg}\sqrt{1-(x/d)^2},
\end{equation}
where, $E^0_{\rm bg}=2\pi e^2n_0d/\kappa$ presents the global minimum of the background potential and $e$ is  the elementary charge. For typical samples, $E^0_{\rm bg}$ is at the order of 4 to 5 eV and is used to normalize energy in our numerical calculations while it is independent of magnetic field strength. The electron density can be obtained by solving Poisson analytically within CSG model without magnetic field as,
\begin{equation}
n(x)=n_0\left( \frac{x-l}{x+l}\right)^{1/2},
\end{equation}
here $n_0$ is the bulk electron density far from the edges and $l$ is the depletion length, defining the width of electron free zone. Since the electrons are repelled by negatively charged gates from the edges, $n_{\rm el}(x)$ is confined to the interval $ |x|<l $.
Magnetic field influences the electron density distribution locally due to Landau quantization. Hence 2DES is separated in two different regions. One with highly degenerate  and the other with approximately vanishing DOS. The region with high DOS acts metal-like and has good screening ability, which is called as compressible. The other region acts as insulator-like, with almost vanishing DOS. This incompressible region has poor screening ability, where the electron density is constant. Whereas, the total potential presents a gradual behavior on opposing edges with opposite sign of their slopes.

\subsection{Self-consistent Thomas - Fermi Approximation (SCTFPA)}
If the self-consistent total (screened) potential $V(\vec{r})$ varies slowly compared to quantum mechanical quantities, it can be added as a constant to Landau Hamiltonian,
\begin{equation}
H=\frac{1}{2m}\left({\hat{p}({\vec{r}})}-\frac{e}{c}{\mathbf{A}(\vec{r}})\right)^2 + V(\vec{r}),
\end{equation}
where $\mathbf{A}(\vec r)$ is vector potential that defines magnetic field
$\mathbf{B}(\vec{r})=\vec{\nabla} \times \mathbf{A}(\vec{r}) $, and $\hat{p}(\vec{r})$
is the momentum operator. We use the Landau (Coulomb) gauge $\mathbf{A}(\vec{r})=(0,B_x,0)$ to solve the Hamiltonian yielding 1D harmonic oscillator wave functions in $x$ and plane waves in $y$ as, 
\begin{equation}
\phi_{n,X}(\vec{r})=L_y^{-1/2}e^{ik_y.y}\psi_{n,X}(x)
\end{equation}
 $L_y$ is the sample length and $k_y$ is the wave-vector in $y$ direction. The function $\psi_{n,X}(x)$ is nothing but the eigenfunction of simple harmonic oscillator, which is centered at $X$, the center coordinate and is given as $X=-\ell^2k_y$, where the magnetic length is $\ell=\sqrt{eB/\hbar}$ being proportional to the width of $n^{\rm th}$ Landau wave-function and cyclotron radius $R_c$.
 
 The energy eigenvalues of this Hamiltonian are $E_{n,X}$ depending on Landau index and the center coordinate. TFA  assumes that electrons are point-like particles, hence, energy eigenvalues  approximates to  
\begin{equation}
E_{n,X}\approx E_n+V(X),
\end{equation}

In Hartree approximation, which also takes into account finite widths of the Landau wave-function, electron density can be found as,
\begin{equation}
n_{\rm el}(x)=\frac{g_s}{2\pi l^2}\sum_{n,X}\int dx f(E_{n,X}(x)-\mu^*)\left| \phi_{n,X}(x)\right|^2,
\end{equation}
where Fermi function is $f(E)=1/(1+\left[\exp(E/k_{\rm B}T)\right] )$, $k_{\rm B}$ being the Boltzmann constant and  $\mu^*$ is electrochemical potential that has a fixed value in equilibrium state. Setting $g_s=2$ neglects spin degeneracy and $\phi_{n,X}(x)$ is the wave function obtained by solving Hamiltonian. A fundamental Density Functional Theory (DFT) approach~\cite{kohnsham}, i.e. TFA~\cite{Asch:book}, reduces the complexity of Hartree approximation in a reasonable manner by letting $\left| \phi_{n,X}(x)\right|^2\approx \delta(x-X)$. It is important to emphasize that, Landau wave-functions have a Gaussian form and the Hamiltonian can be reduced to a simple harmonic oscillator in one dimension. Hence, satisfies the minimum uncertainty principle, therefore as a first order approach TFA finds its rationale on firm grounds.  As a consequence, electron density reduces to,

\begin{equation}
n_{\rm el}(x)=\int dE D_{B}(E) f(E+V(x)-\mu^*),\label{Eq:elecdensity_TFA}
\end{equation}
where $D_{B}(E)$ is the DOS in the presence of a perpendicular magnetic field. Considering an ideal (without scattering broadening) $D_{B}(E)$ is the Landau density of states,
\begin{equation}
D_{B}(E)=D_{\rm L}(E)=\frac{g_s}{2\pi \ell^2}\sum_{n=0}^{\infty}\delta(E-E_n).
\end{equation}
Recall that, Landau DOS does not include temperature effects, which directly affects occupation of energy levels. Consequently, the highest occupied level is not equal to $E_{\rm F}$ as at $T=0$, instead equates to chemical potential, which is a thermodynamic quantity. Remember that in equilibrium (without an external current imposed), chemical potential $\mu$ is equal to electrochemical potential $\mu^*$. Hence, we need to consider the effect of chemical potential on DOS which is then called thermodynamic density of states (TDOS), $ D_{\rm T}(E)=dn_{\rm el}/d\mu^*$. For bare Landau quantized density of states TDOS reads,
\begin{equation}
D_{\rm T}(E)=\frac{g_s}{2\pi l^2}\sum_{n=0}^{\infty}\frac{\beta}{4\cosh^2(\beta[E_n-\mu^*]/2)}\label{Eq:TDOS_B},
\end{equation}
where $\beta=1/k_{\rm B}T$.

In our approach we obtain the electron density distribution, following the above mentioned works by Gerhardts group, starting with a situation without electrons at $T=0$. In this step we assume that the system is in a natural state, i.e. $n_0=n_{\rm e}$, which in turn fixes the $E_{\rm F}$, i.e. the total number of electrons is constant even at finite temperature, $B$ field and imposed external current. Due to negatively charged gate electrodes electrons are depleted from the edges. Thus, $n_{\rm el}(x)$ has a spatial dependence different from the spatially homogeneous donor distribution. Note that assuming translational invariance together with the Thomas-Fermi screening theory, enables us to calculate the screened potential by,
\begin{equation}
V_{\rm Scr}(\vec{k})=[\kappa(1+\frac{|\vec{k^2_F}|}{|\vec{k}^2|})]^{-1}V_{\rm Ext}(\vec{k})
\end{equation}
where $|\vec{k}{_F|}$ is the Fermi and $|\vec{k}|$ is the wave-vector of an electron in $xy-$ plane. Essentially, the expression $(1+\frac{|\vec{k^2_F}|}{|\vec{k^2}|})$ is the dielectric function $\varepsilon(k)$. Note that, at finite temperatures $|\vec{k_F}|$ is replaced by $|k_0|=\sqrt{4\pi e^2 \partial n_0/\partial\mu^*}$, and $\partial n_0/\partial\mu^*$ is nothing but the TDOS. Recall that due to charge neutrality $n_0=n_{\rm el}$. We find it important to remind the reader that in case of equilibrium, namely if no external current is imposed $\mu^*=\mu$, for a homogeneous system.

The SCTFPA uses screened potential energy as an initial condition which is calculated only for donors at $T=0$ and $B=0$. In the next numerical step, screened potential and electron density distributions are calculated at $B>0$ and relatively high temperatures, namely $\hbar \omega_c/ kT\ll1$. Since, the Fermi distribution becomes smoother in energy by temperature effects, it is rather easy task to obtain $V(x)$ and $n_{\rm el}(x)$ while any divergent behavior that can pop up due to singularities in Eq.~(\ref{Eq:elecdensity_TFA}), is eliminated. Starting at a relatively high $T$, Newton-Raphson iteration scheme is used and the temperature is lowered step-by-step to obtain electron density distribution at targeted $T$ and $B$. In each iteration step difference between the previous step is calculated for the fixed $\mu^*$ and if the difference is less then $10^{-7}$, the temperature is lowered one step further. 

\subsection{Corrections to 2D DOS}
In realistic systems DOS broadens due to impurity scattering. This means that the $\delta$ form of bare Landau DOS can be expressed in terms of the spectral function, $A_{n,\sigma}(E)$, where $n$ is the Landau index and $\sigma$ is the spin degree of freedom which we neglect as,
\begin{equation}
D_{\rm L}(E)=\frac{g_s}{2\pi \ell^2}\sum_{n=0}^{\infty}\delta(E-E_n)=\frac{g_s}{2\pi \ell^2}\sum_{n=0}^{\infty}A^{\rm L}_{n}(E).
\end{equation}
The DOS of such an idealized QH system does not always yield reasonable results when considering real samples, in particular when the mobility of regarding sample is low (high scattering due to impurities) and/or relatively high at temperatures $T\approx 1-10$ K. Therefore, we also consider two different DOS models regarding spectral functions. 

First one has a Gaussian form,
\begin{equation}
A_{n}^G(E)=(2\pi\Gamma_{n}^2 )^{-1/2}  \exp\left(-\frac{ 1}{2} \left[ \frac{(E-E_{n})}{\Gamma_{n}} \right] ^2 \right),
\end{equation}
where $\Gamma_{n}$, is the half-width-full-maximum of Gaussian DOS. However, such a DOS lacks an energy gap without assuming localization~\cite{Kramer03:172} (i.e. mobility gap). Since the tails of Gaussian distribution approximates to zero only at infinity there will be a small but finite overlap between all energy levels. In an improved model we utilize Self-Consistent Born Approximation (SCBA) to overcome this discrepancy, which yields semi elliptical DOS. Within the SCBA one obtains mutually an energy gap in density of states. SCBA spectral function is obtained as~\cite{Ando74:959,Ando82:437},

\begin{equation}
A_{n}^{SCBA}(E)=\left( \frac{1}{\pi\Gamma_{n}}\right)  \exp{\left(1-\left[ \frac{(E-E_{n})}{2\Gamma_{n}} \right] ^2 \right)^{1/2}}.
\end{equation}

In consequent subsections, we calculate local electron densities and longitudinal conductivity spatially, under above discussed approximations. 
\begin{figure}
	\centering{
		\includegraphics[scale=1.]{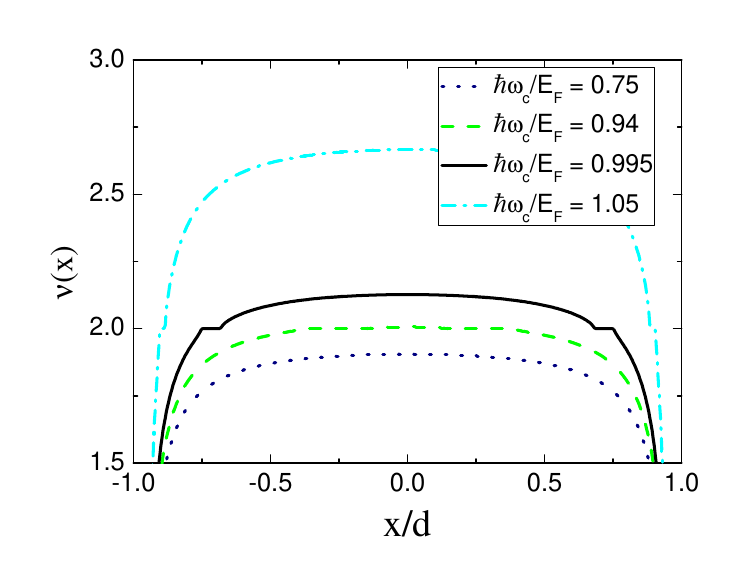}
		\includegraphics[scale=1.]{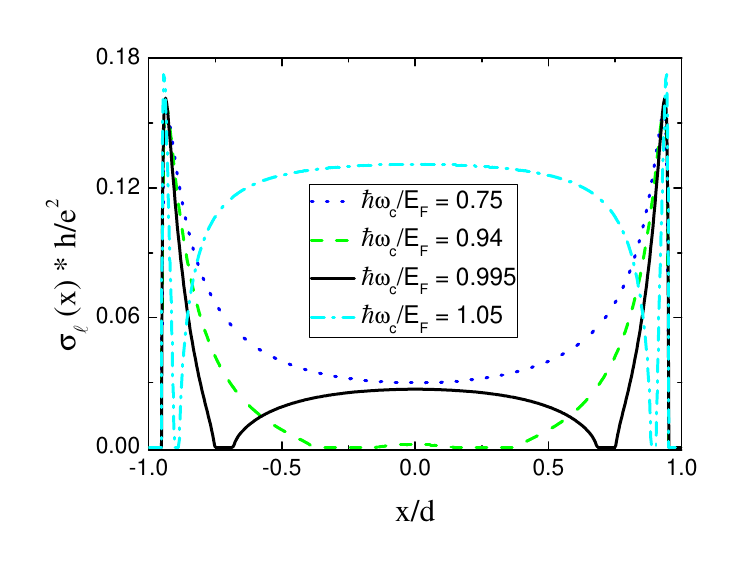}}
	\centering
	\caption {Self-consistent numerical results calculated at four characteristic $B$, within TFPA and using a Gaussian DOS profile. (a) The local filling factor profile as a function of normalized coordinate $x/d$, assuming $d=1.5 \quad \mu$m. The depletion length is taken to be $d/10$, whereas donor density is set to be $n_0=4\times 10^{-15}$ m$^{-2}$. (b) Position dependent longitudinal conductivities. We impose that level broadening is independent of Landau index $n$, $\Gamma/E_{\rm bg}^0= \gamma{=0.025}$, where the default temperature is set to $ T=2$ K . }
	\label{Fig:1}
\end{figure}
\subsection{Position dependent density profiles and Transport coefficients}
Being equipped with self-consistent numerical scheme to calculate electron density profile $n_{\rm el}(x)$, within the SCTFPA one can straight forwardly obtain the local filling factors $\nu(x)$ for a given $B$ field strength $\nu(x)=2\pi\ell^2n_{\rm el}(x)$. In Fig.~\ref{Fig:1}a, we show local filling factors as a function of position, which is normalized to the sample half width $x/d$. The cyclotron energy $\hbar \omega_c$ depends only on the strength of $B$, hence scaling $\hbar \omega_c$ by the constant $E_{\rm F}$ enables us to compare our results with the experimental ones directly. Note that, $2/\bar{\nu}=B$ and $2\hbar \omega_c/E_{\rm F}=\bar{\nu}$, where $\bar{\nu}$ is the average filling factor which differs from that of calculated at the center of sample $\nu(x=0)$ considering finite and rather narrow samples. Due to the fact that electrons are depleted from the edges this different is $\nu(0)-\bar{\nu}=\Delta \nu$. In the case of $l\ll2d$, $\Delta \nu\rightarrow 0$.

In Fig.~\ref{Fig:1}a we present filling factor distributions calculated for four characteristic $B$ fields. At the highest $B$ ($\bar{\nu}<2$) 2DES is completely compressible (dotted lines), since at each location there are available states at electrochemical potential, $\mu^*=\mu(x)-V(x)$. Lowering $B$ ($\bar{\nu}\lesssim 2$) results in a situation where bulk of 2DES becomes incompressible, namely approximately no states are available (broken lines) at the bulk to screen total potential. At the next lower field with $\bar{\nu}\gtrsim 2$ we observe that two incompressible strips are formed at the opposing edges with constant electron density, whereas the rest of 2DES is compressible (thick line). Lower most $B$ presents very narrow incompressible strips in the proximity of edges, which can be considered as compressible due to thermodynamic reasons as discussed below.

It is crucially important to clarify the expression ``very narrow incompressible strip":  As discussed in the Introduction, compressibility is a thermodynamic quantity. Once the dimensional scales of the incompressible strip becomes comparable with thermodynamic lengths, e.g. thermal or Fermi wavelength, one has to take care of previously made assumptions on local equilibrium. To define thermodynamically an incompressible strip in a consistent manner, it is obligatory to compare the widths of incompressible strips $w_{n}$ ($n$ being the integer level index of $\nu(x)=n$) with thermodynamical length $L_{T}$ scale(s). For vanishing temperature and without $B$ field, $L_{T}$ simply equates to $\lambda_{\rm F}$, the Fermi wavelength. However, at low and finite temperatures ($10\lesssim T \lesssim 20$ K, for GaAs samples) it is reasonable to take $L_{T}\simeq \lambda_{\rm th}=h/(\sqrt{2\pi m^*kT})$, where $\lambda_{\rm th}$ is the (de Broglie) thermal wavelength. At higher temperatures, i.e if $\lambda_{\rm F}\gg\lambda_{\rm th}$, the 2DES is considered as Maxwell-Boltzmann classical ideal gas. A back of envelope calculation, considering experimental values of $m^*$ and $T$, one obtains $\lambda_{\rm th}\gtrsim 30-50$ nm. Hence, once the condition $w_n\lesssim\lambda_{\rm th}$ is satisfied, the strip can no longer be well defined as incompressible, thermodynamically. Such a strip is then called evanescent~\cite{Siddiki10:67010}. The existence of these evanescent incompressible strips are confirmed and investigated also by experiments~\cite{Sailer10:113033,Kendirlik13:3133}. Moreover, Salman et. at reports that, it is also possible to investigate the effects of evanescent incompressible strips in the case of fractional quantized Hall effects~\cite{Salman13:203}.

The next objective is to obtain local longitudinal conductivities $\sigma_{l}(x)$ for desired $B$ fields, to link local capacitances, as measured in experiments. Transport quantities can be obtained in a self-standing manner by assuming local equilibrium within the linear response regime and Drude formalism~\cite{Guven03:115327,siddiki2004}. Essentially, this approach finds its firm grounds on Born-Oppenheimer approximation assuming that, although the electronic levels are quantized, the displacement (drift) experienced by the center of mass (CM) can still be described by classical equations of motion. In our case, the quantized electronic levels are the Landau orbits and the center coordinate ($X$) replaces CM. Then one can reliably describe drift of electrons within the Drude model, driven by the external electric field $\mathbf{E}(\vec{r})$. Below we briefly re-introduce the local equilibrium model developed by Gerhartds and co-workers. Higher order contributions to this approach is discussed in the literature by Champel group~\cite{Champel08:124302} utilizing Greens function formalism considering equilibrium properties and dissipation of the system which we investigate here. In their work, they justify our approach of local equilibrium assumption. Rationale stems from the condition than in linear response regime, density-density and current-current correlations are sufficiently small and can be neglected.  

Applying a potential (energy)  difference between source and drain contacts in the linear response regime with local equilibrium, $eV_{\rm SD}\ll\hbar\omega_{c}$, the electrochemical potential depends on spatial coordinates $\mathbf{\mu}^*(\vec{r})$. The Electric field can then be described as $\mathbf{E}$$(\vec{r})$$=\mathbf{\nabla}$$\mathbf{\mu}$$^*/e$, driving the imposed dissipative total current $I=\int\mathbf{j}(\vec{r})d
\vec{r}$. We remind that our model already assumes translational invariance in the transport direction $y$, hence $E_y(x)$ is constant equating $E_y^0$ and the transverse field $E_x(x)$ is the Hall field, i.e. $\mathbf{E}(\vec{r})=(E_{\rm H}, E_y^0,0)$. The components of current density $\mathbf{j}(\vec{r})$ are given by the local Ohm's relation via,
\begin{equation}
\mathbf{j}(\vec{r})=\mathbf{E}(\vec{r})\hat{\sigma}(\vec{r}),
\end{equation}
$\hat{\sigma}(\vec{r})$ is the two by two conductivity tensor where $y$ dependence drops,

\begin{equation}
\hat{\sigma}(x)=\left[
\begin{array}{c} 
 \sigma_{xx}(x)   \quad  \sigma_{xy}(x)  \\
	-\sigma_{yx}(x)   \quad  \sigma_{yy}(x)  \\ 
\end{array}
\right].
\end{equation}
Assuming local equilibrium (i.e. negligible density-density and current-current correlations) is well justified to write $\hat{\sigma}(x)=\sigma(n_{\rm el}(x))$. To keep the notation consistent with the existing literature we set $\sigma_{xx}(x)=\sigma_{yy}(x)=\sigma_{l}(x)$ and $\sigma_{xy}=-\sigma_{yx}(x)=\sigma_{\rm H}(x)$. Calculating components $j_x(x)=0$ and $E_y(x)=E_y^0$ are rather straightforward due to translational invariance and equation of continuity. Consequently,
\begin{equation}
	j_y(x)=E_y^0/\rho_{l}(x), \quad E_x(x)=(\rho_{\rm H}(x)E_y^0)/\rho_l(x),
\end{equation}
where $\rho_{xx}(x)=\rho_{l}(x)$, $\rho_{xy}(x)=\rho_{\rm H}(x)$. Note that in two dimensions, resistivity tensor is given by
 \begin{equation}
 \hat{\rho}(x)=
\frac{\sigma_{l}(x)\sigma_{\rm H}(x)}{\sigma^2_{l}(x)+\sigma^2_{\rm H}(x)}
\left[
 \begin{array}{c} 
 \sigma_{l}(x)  \quad    \sigma_{\rm H}(x)  \\
 -\sigma_{\rm H}(x)   \quad     \sigma_{l}(x)  \\ 
 \end{array}
 \right].
 \end{equation}
 
After re-introducing the self-consistent calculation scheme to obtain local electron densities and conductivities, in this part we discuss how to include thermodynamic effects to our model. As mentioned before, incompressibility and capacitance obtained from conductivity are both thermodynamical quantities. Hence, one should take into account temperature and (electro-)chemical potential, while calculating local conductivities. We do this by a simple, however, fundamental interpolation method, essentially averaging local conductivities over the thermal wavelength $\lambda_{\rm th}$:

\begin{equation}
\hat{\bar{\sigma}}(x)=\int_{-\lambda_{\rm th}}^{\lambda_{\rm th}}d\zeta\hat{\sigma}(x+\zeta).
\end{equation}   
Using above discussed averaging scheme, removes artificial singularities that may pop-up in local conductivities. Below we use this interpolated quantities to describe global experimental observables. 
 
At this point, calculating local transport coefficients and global measurable quantities, such as $R_L$ and $R_{\rm H}$, are simplified to obtain local electron density. The total current is expressed as $I=\int_{-d}^{d}dxj_y(x)$ enabling us to calculate constant electric field in current direction,
\begin{equation}
	E_y^0= I\left[\int_{-d}^{d}dx\frac{1}{\rho_l(x)}\right],
\end{equation}
 transverse (Hall) direction then is obtained as,
 \begin{equation}
 V_{\rm H}=\int_{-d}^ddxE_x(x)=E_y^0\int_{-d}^ddx\frac{\rho_{\rm H}(x)}{\rho_l(x)}.
 \end{equation}
Global resistances are obtained for a square sample yields,
 \begin{equation}
R_{\rm H}=\frac{V_{\rm H}}{I} \quad \mathtt{ and } \quad R_l=\frac{2dE_0^y}{I}.
\end{equation} 
 
The longitudinal conductivity can also be expressed in terms of spectral and Fermi functions~\cite{Gerhardts75:285},
\begin{equation}
\sigma_{l}=\frac{e^{2}g_{s}}{h}\int_{-\infty}^{\infty}dE\left[-\frac{df}{dE}\right]\sum_{n=0}^{\infty}\left(n+\frac{1}{2} \right) \left[ \sqrt{\pi} \Gamma_{n} A_{n}(E)\right]^2,
\end{equation}  
which allows us to embed the thermal and scattering broadening effects when calculating local conductivities. 

In the following we still assume spin-less electrons for the Gaussian spectral function and neglect effects of the level index $n$ in level broadening  parameter, i.e. $\Gamma_{n}=\Gamma$. However, for the SCBA we take into account level broadening also depending on level index, which is more realistic compared to the Gaussian model.

Fig.~\ref{Fig:1}b, shows the corresponding local longitudinal conductivities for the filling factor distributions presented in Fig.~\ref{Fig:1}a. At the highest $B$ field, since all the 2DES is compressible conductivity just follows the density distribution up to a factor. Once the bulk becomes incompressible, $\hbar\omega_c/E_{\rm F}\simeq 0.995$, $\sigma_{l}$ approximates to zero, whereas edges present finite conductivity. Lowering $\hbar\omega_c/E_{\rm F}$  results in a situation where two well developed incompressible strips ($w_{2}>\lambda_{\rm th}$) reside close to both edges, and conductivity becomes exponentially small, however not exactly zero due to finite temperature. In the case of evanescent edge incompressible strips ($w_2\leq \lambda_{\rm th}$), $\sigma_l$ has two local minima exactly at the positions of these strips. Whereas, rest of the sample acts as metal-like, with relatively high conductivity.

\begin{figure}[ht]	
	\centering
	\includegraphics[scale=.4]{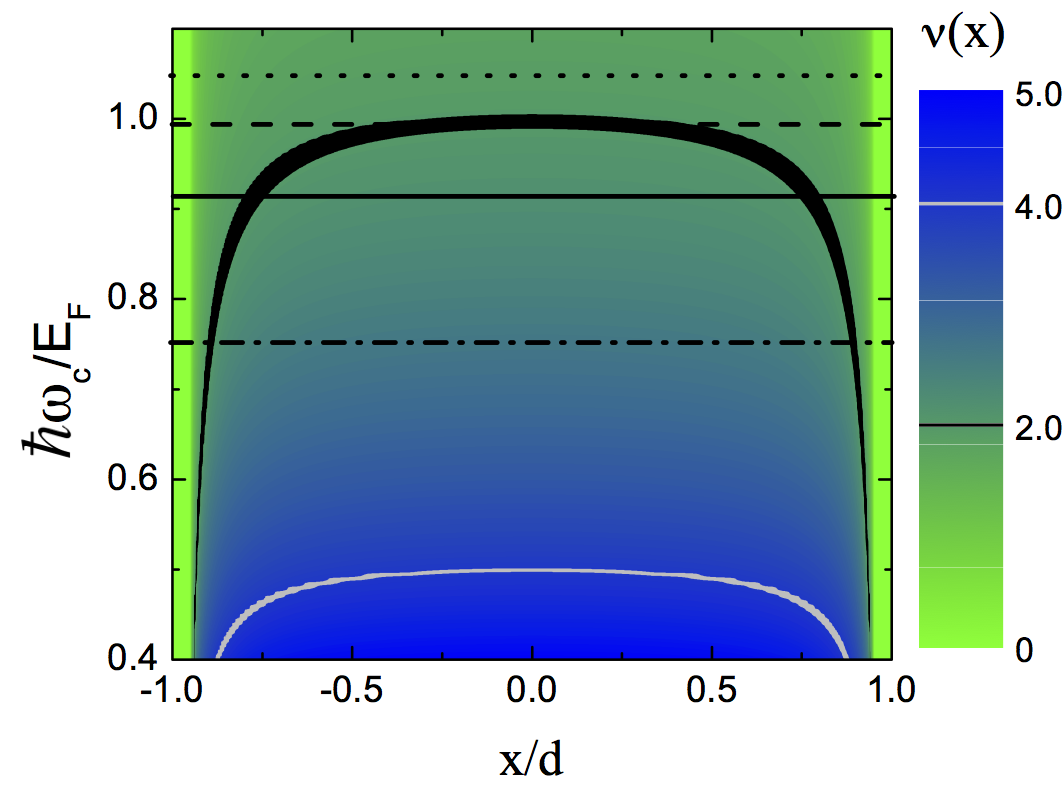}
	\includegraphics[scale=.3]{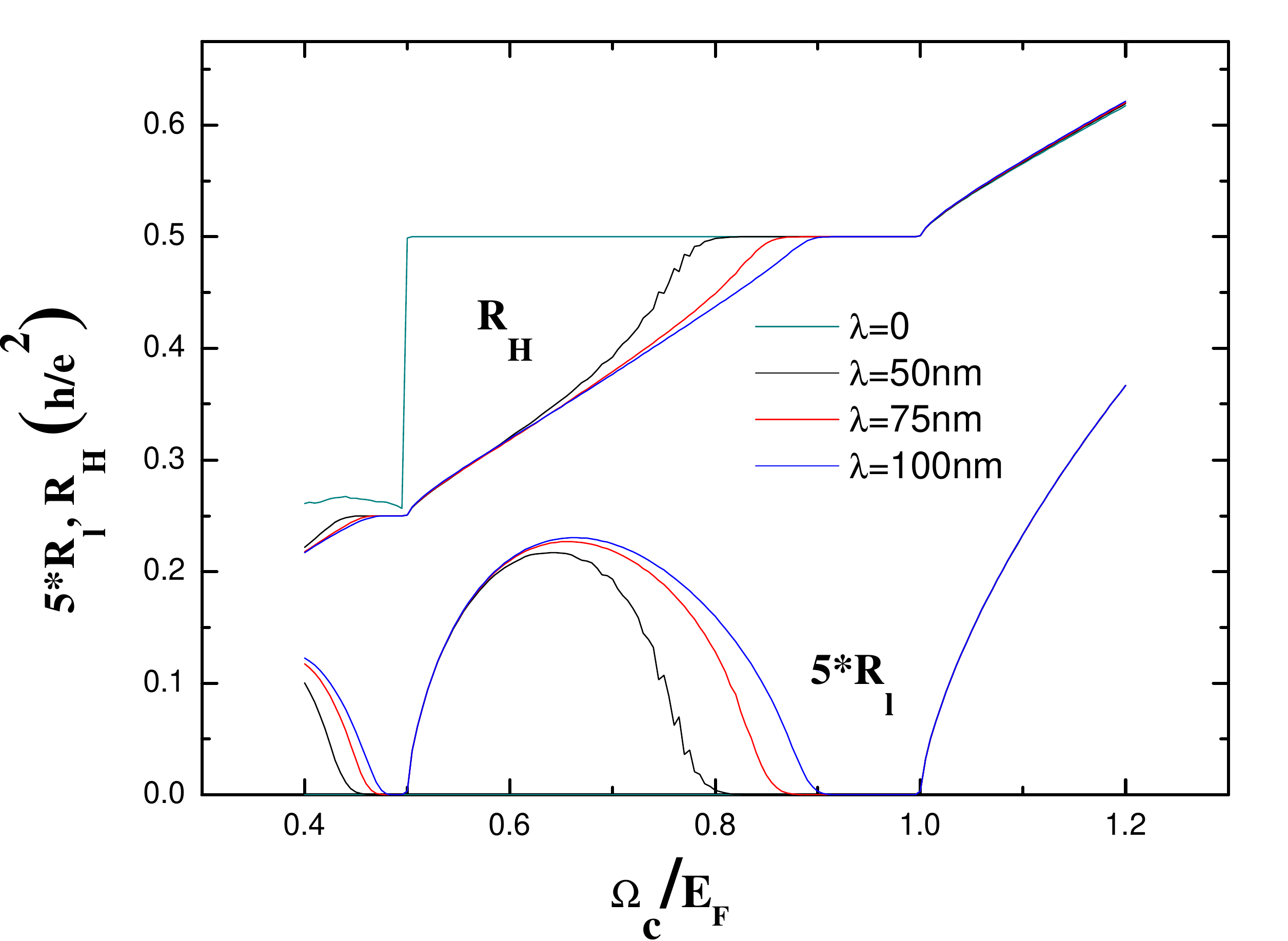}
	\caption{(a) Gray (color) scaled visualization of $B$ scan in the interval $0.4<\hbar\omega_c/E_{\rm F}<1.1$ versus the position dependent filling factors, at default level broadening and temperature. The horizontal lines correspond to $B$ values shown in Fig.~\ref{Fig:1}. Dark (black) zones are incompressible with $\nu(x)=2$, whereas light (yellow) regions indicate  $\nu(x)=4$. (b) Corresponding global resistances as a function of $B$ in units of $\hbar\omega_c/E_{\rm F}=\Omega_c/E_{\rm F}$, where $\lambda_{\rm th}=\lambda$ takes four typical values between 0-100 nm. If the effect of thermal wavelength on local conductivities is neglected (thin line, $\lambda=0$ nm), longitudinal resistance vanishes once an incompressible strip is formed meanwhile $R_H$ is just a constant $\frac{h}{2e^2}$. For finite $\lambda$, one obtains realistic resistances depending on level broadening and finite temperature. All parameters are same as in Fig.~\ref{Fig:1}}  
	\label{Fig:2}
\end{figure}

The spatial evolution of incompressible/compressible regions as a function of $x/d$ and $B$ is shown in Fig.~\ref{Fig:2}a. The gray (color) scale denotes the compressible, whereas dark and light areas correspond to incompressible regions with local values $\nu(x)=2$ and $\nu(x)=4$, respectively. It is clearly seen that when the lowest Landau level is partially occupied, i.e $\bar{\nu}<2$, screening is good and 2DES acts as compressible. System becomes locally incompressible starting from center of the sample, while highest electron density occurs at $x=0$, $\hbar\omega_c/E_{F}\simeq 1$. In the close proximity of this field strength, besides the central region being incompressible, all the rest of 2DES is compressible. The wide bulk incompressible region separates into two strips (see Fig.~\ref{Fig:1}a) which moves towards edges once $B$ is lowered. The incompressible strips with local $\nu(x)=2$ are emphasized by dark (black) in gray (color) scale.
There are three compressible regions decoupled from each other by the incompressible strip(s) residing between them. For the $B$ interval  $0.8\lesssim\hbar\omega_c/E_{F}\lesssim 0.96$, the screening properties can be summarized as follows: We begin our discussion starting from the left most edge ($x=-d$): (\textit{i}) $-d<x<-l$, interval is the electron depleted region in which screening does not takes place. However, the confinement potential is homogeneously damped due to the finite dielectric constant $\kappa$ of the heterostructure. (\textit{ii}) $-l<x<|w^{\rm L}_2|$, 2DES is compressible in the region where $w^{\rm L}_2$ is the left most starting location of the incompressible strip with $\nu(x)=2$. (\textit{iii}) The first poor screening strip is observed in the interval $w^{\rm L}_2<x<w^{\rm R}_2$ and the width of incompressible region is $w_2=|w^{\rm L}_2|+|w^{\rm R}_2|$. This incompressible strip decouples both thermodynamically and electrically the left most compressible region from the bulk with $\nu(x)\gtrsim2$. (\textit{iv}) The second incompressible strip on right-hand-side $(x/d>0$ again decouples bulk and the compressible region at the right edge. In short, we can conclude that if an incompressible strip or region resides somewhere in the sample, left and right compressible regions are decoupled yielding quantized Hall resistance. In addition, since the longitudinal resistivity vanishes along the incompressible strip due to lack of scattering, the longitudinal resistance approximates to zero, decaying exponentially depending on temperature.

In the next Section, we summarize the experimental techniques that measures local capacitances obtained via local conductivities. 
\section{Experimental Background}{\label{Sec.3}}
Recent experiments focusing on the edge properties of quantum Hall effect brought interest on the capacitance of compressible and incompressible regions within the 2DES subject to high perpendicular $B$ fields. In a relatively recently developed experimental setup utilizes dynamic scanning capacitance microscope (DSCM) operating between 1.9 K and room temperature. An atomic force microscope (AFM) tip scans the surface of GaAs/AlGaAs heterostructure, where the spatially measured signal is proportional to the local conductance and capacitance of the 2DES beneath surface. During measurements the system works under 2 mbar atmospheric pressure in the non-contact mode. Magnetic field is scanned between 0 and 12 T. 
Electrical measurements are performed by two lock-in amplifiers by which output signals can be separated as imaginary and real parts. There are essentially two important frequencies, first one is the frequency of oscillating AFM tip ($\omega$) and the second one is the resonance frequency of lock - in amplifier with the output signal of AFM tip ($\omega_{rf}$). These measured signals are proportional to tip-sample conductance, $G_{ts}$ via,
\begin{equation}
G_{ts,\omega}\approx\frac{dG_{ts}}{dC_{\rm ts}}C_{\rm ts,\omega}=\omega_{rf} C_{ts,\omega} (\omega 2R_S C_{ts}+i)\label{Eq:gts}
\end{equation}
where $R_S$ is resistance of 2DES and $C_{ts}$ is the total tip-sample capacitance and equals to sum of classical and quantum capacitances which are related to compressibility of 2DES~\cite{Suddards12:083015}. Classical (geometric) capacitance $C_{\rm geo}$ is considered as serially connected two pairs of parallel plate capacitors. One of these capacitances is between metallic surface and the tip $C_1=z/\varepsilon_0$, where $z$ is the distance between surface-tip and $\varepsilon_0$ being dielectric constant of vacuum. The second one is mutual electrostatic capacitance between 2DES and metallic surface $C_2=d/(\kappa\varepsilon_0)$. As expressed in Eq.~\ref{Eq:1}, $C^{-1}_{\rm geo}=(1/C_1+1/C_2)$. Quantum capacitance $C_3=C_{\rm q}$ is also serially connected to classical one and is proportional to DOS of 2DES. Then total capacitance reads,
\begin{equation}
C_{ ts}=\left( \frac{1}{C_1}+\frac{1}{C_2}+\frac{1}{C_3} \right)^{-1},
\end{equation}

\begin{equation}
C_{ts}=A\left( \frac{z}{\varepsilon_{0}}+\frac{d}{\kappa\varepsilon_{0}}+\frac{1}{e^2D_T}\right)^{-1} .
\end{equation}
Calculating $G_{ts}$ from Eq.~\ref{Eq:gts} requires the resistance between ground and sample,
\begin{equation}
R_S= \frac{\ln(r_2/r_1)}{2\pi\sigma_{l}^{(2DAu)}},
\end{equation}
which is extracted from experiments performed on a thin (approximately 2D) gold film, by measuring the 2D conductivity. Here, $r_1$ is the radius of AFM tip, $r_2$ defines the circular boundaries of 2D gold film and $\sigma_{l}^{(2DAu)}$ stands for the longitudinal conductivity. Applying this model to 2DES one can obtain the experimental value of $R_S$. As described above, since longitudinal conductivity can be calculated numerically within the Thomas-Fermi-Poisson approximation, as the base of our approach, in the following Sections we compare experimental results with our findings.

Finally we need to calculate $C_{ts}$, around integer filling factors.  We obtain the real and imaginary parts of $G_{ts}$ neglecting constant terms enables us to simplify numerical calculation without touching the essence of our approach, while constant terms already drop due to differentiation. The total geometric capacitance is symbolized by $C_{\rm geo}$, whereas the quantum counterpart is denoted by $C_{\rm q}$. Real part of $G_{ts}$ is,
\begin{equation}
\Re\left\lbrace G_{ts}\right\rbrace =\left(\frac{1}{C_{\rm geo}}+\frac{1}{C_{\rm q}} \right)^{-2} R_S,
\end{equation}
\begin{figure}
	\centering
	\includegraphics[scale=.5]{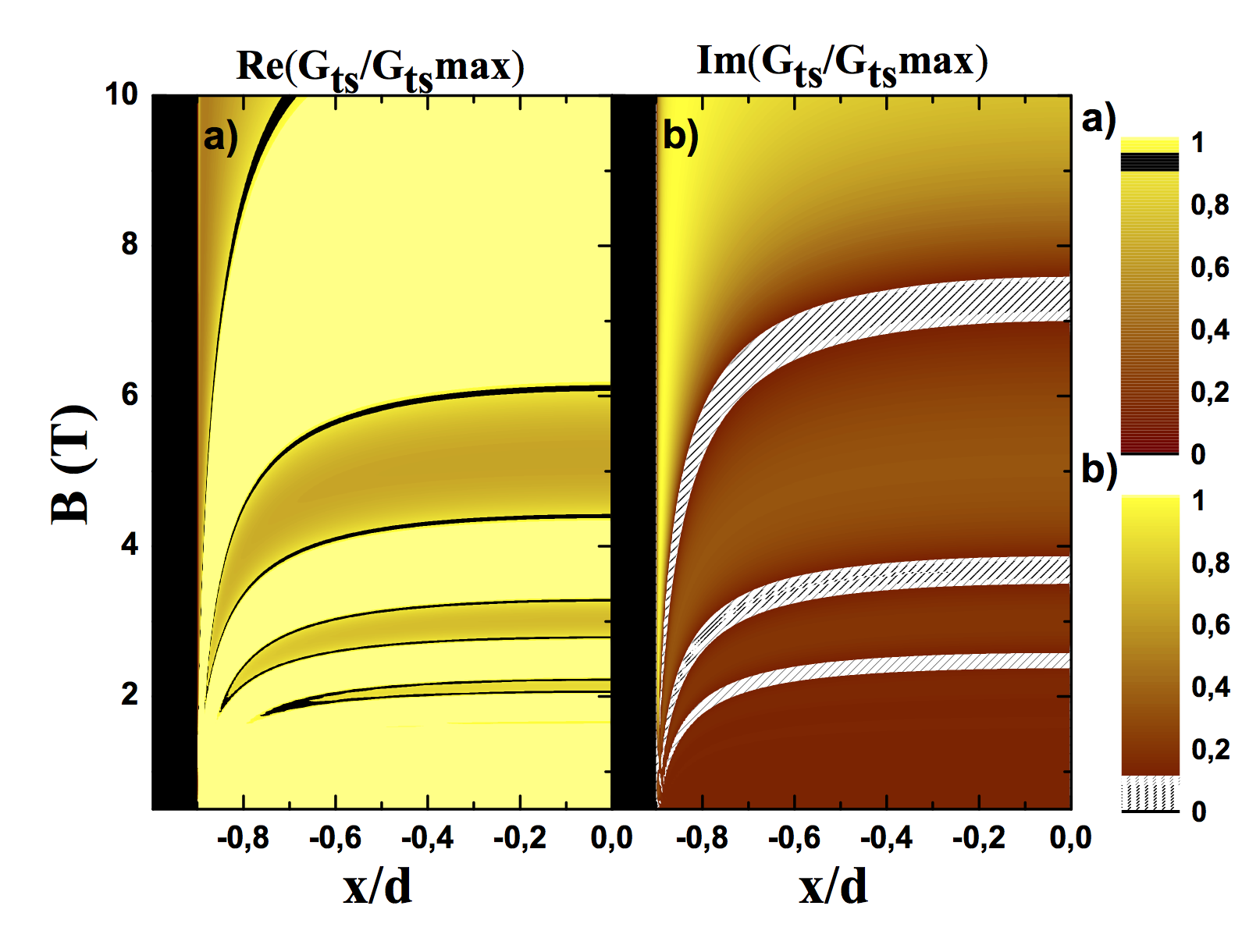}
	\caption {Numerical calculations of tip-sample conductance $G_{ts}$, where Gaussian DOS is assumed and local conductivities are obtained within TFPA. (a) Real part of $G_{ts}$ in gray (color) scale as a function of position $x/d$ and magnetic field $B$. To directly compare our results with the experiments, we back normalized field strength to Tesla, which varies in the interval [0, 10] T. Black lines contouring light (yellow) regions depict the transition from compressible to incompressible.  Light (yellow) regions indicate that $\Re G_{ts}$ is dominated by the geometrical capacitance, while both $C_{\rm q}$ and $R_S$ approximates to zero. (b) The imaginary part of $G_{ts}$. Patterned white regions present accurately the evolution of incompressible strips, since in this situation only the quantum capacitance is the dominating parameter. Calculations are performed at default temperature and by setting $\gamma=0.1$ results in a wide level broadening, therefore smears out narrow incompressible strips efficiently.}
	\label{Fig:3}
\end{figure}
which can be written in an explicit form as
\begin{equation}
\Re \left\lbrace G_{ts}\right\rbrace = C_{\rm geo}^2\left(\frac{C_{\rm q}}{C_{\rm geo} + C_{\rm q}} \right)^{-2} R_S,
\end{equation}
where $C_{\rm geo}$ is constant while varying magnetic field. However, $C_{\rm q}$ strongly depends on $D_B(E)$ as function of magnetic field and, in particular at integer filling factors $C_{\rm q}$ approximates zero, due to $D_B(E)\approxeq0$. Therefore,  $C_{\rm geo}$ dominates $G_{ts}$. Similar considerations can be made for the imaginary part,
\begin{equation}
\Im \left\lbrace G_{ts}\right\rbrace = \left(\frac{C_{\rm q}}{C_{\rm geo} + C_{\rm q}} \right)^{-1}.
\end{equation}
Realize that, while calculating $\Im \left\lbrace G_{ts}\right\rbrace$, $C_{\rm q}$ term in denominator becomes sufficiently small if $\nu(x)\approx n$, hence one can neglect it. Then imaginary part predominantly depends on $C_{\rm q}$ term.

In light of above discussions on the local conductances and capacitances, next we present our findings considering different DOS forms, level broadening widths and thermal wavelengths.  

\section{Numerical Results and Discussion}
In this Section, first we point out the effect of level broadening on conductivities by comparing highly disordered sample resulting in wide level broadening and almost an ideal 2DES within the Gaussian DOS model. Next we present our numerical findings obtained within the SCBA model and show that the latter form of DOS yields more realistic results compared to experimental ones. In the last step, we also present the global quantity $R_l$ obtained by the local Ohm's relation.
\begin{figure}
	\centering
	\includegraphics[scale=.5]{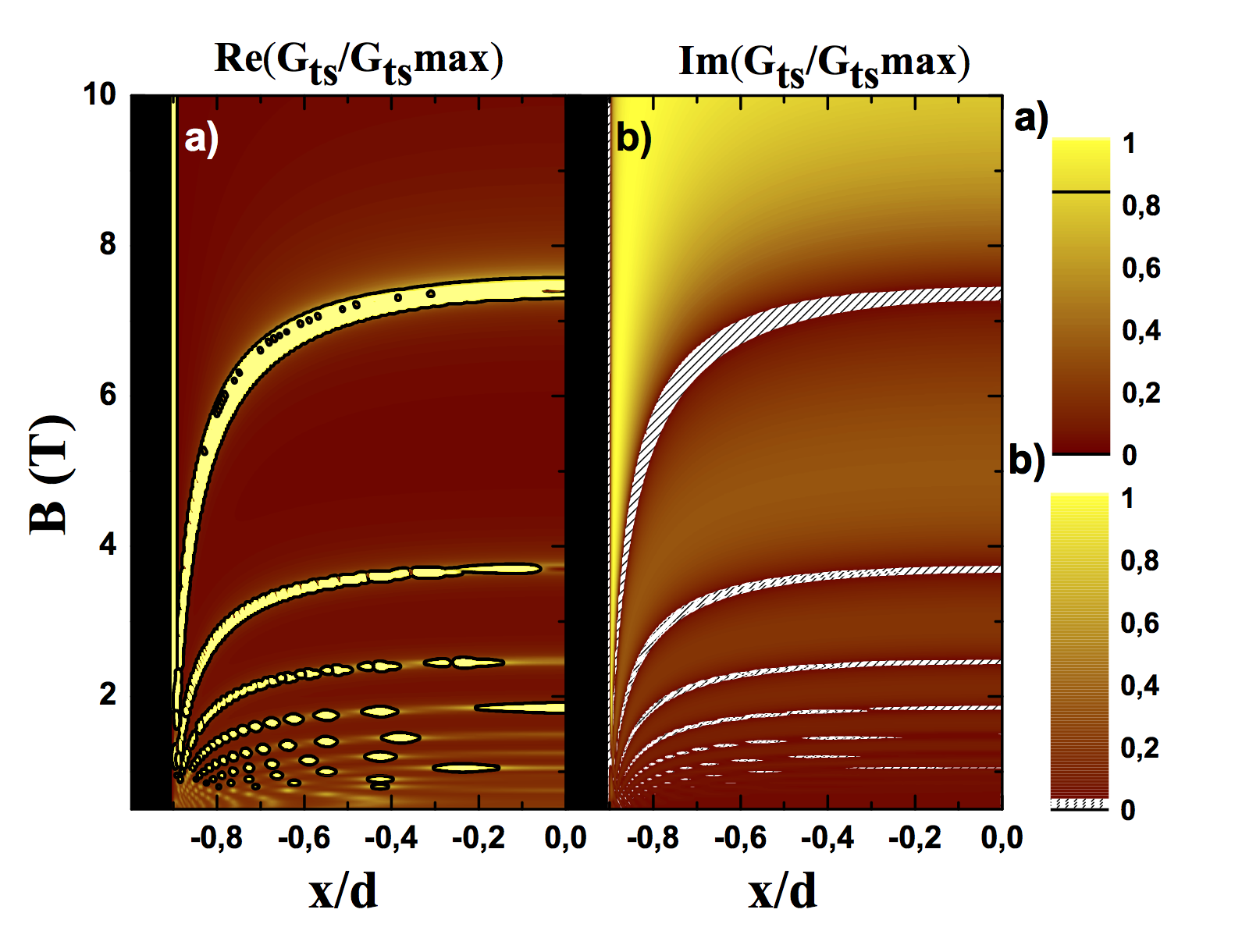}
	\caption {Same as Fig.~\ref{Fig:3}, however, assuming a low-impurity sample by setting $ \gamma{=0.025}$ at default temperature. Considering narrow broadened Gaussian DOS leads co-existence of many incompressible regions, which has the similar effect on global resistances when thermal length $\lambda=0$, as shown in Fig.\ref{Fig:1}b.} 
	\label{Fig:4}
\end{figure}
Fig.\ref{Fig:3} plots both the imaginary and real parts of tip-sample conductance normalized by the maximum value of $G_{ts}$. The real part presents large light areas, indicating that these regions are highly conductive, which means that the geometrical capacitance is dominating. This behavior points that, mentioned areas are compressible. Darker areas contoured by black lines are relatively less conducting, which we expect such a situation for incompressible regions. The imaginary part is shown in Fig.\ref{Fig:3}b, which is dominated by the quantum capacitance. Patterned white regions present approximately zero conduction and clearly indicates incompressible areas, while total capacitance  reaches its maximum value where DOS vanishes. We use the Gaussian DOS assuming a relatively wide level broadening to mimic a low-mobility sample. Due to high number of impurities and finite overlap between energy levels, narrow incompressible strips are washed out. To clarify the effect of level broadening within the Gaussian model we show same quantities in Fig.\ref{Fig:4}, however, here we consider a high-mobility sample by setting the level broadening to be sufficiently narrow, i.e. $\gamma=0.025$. In this situation, the overlap between levels are reduced and narrow incompressible strips can survive. This results in an effective decoupling of compressible regions by the incompressible strips. It is worthwhile to emphasize that the low conducting areas (darker regions in Fig.\ref{Fig:4}a and light-shaded areas in Fig.\ref{Fig:4}b) seems to be almost at the same $B$ intervals. However, realize that the high conducting regions are slightly shifted to stronger $B$, compared to low conducting incompressible strip $B$ strengths. This is another proof that the compressible regions are decoupled by the incompressible ones.       
\begin{figure}
\centering
\includegraphics[scale=.5]{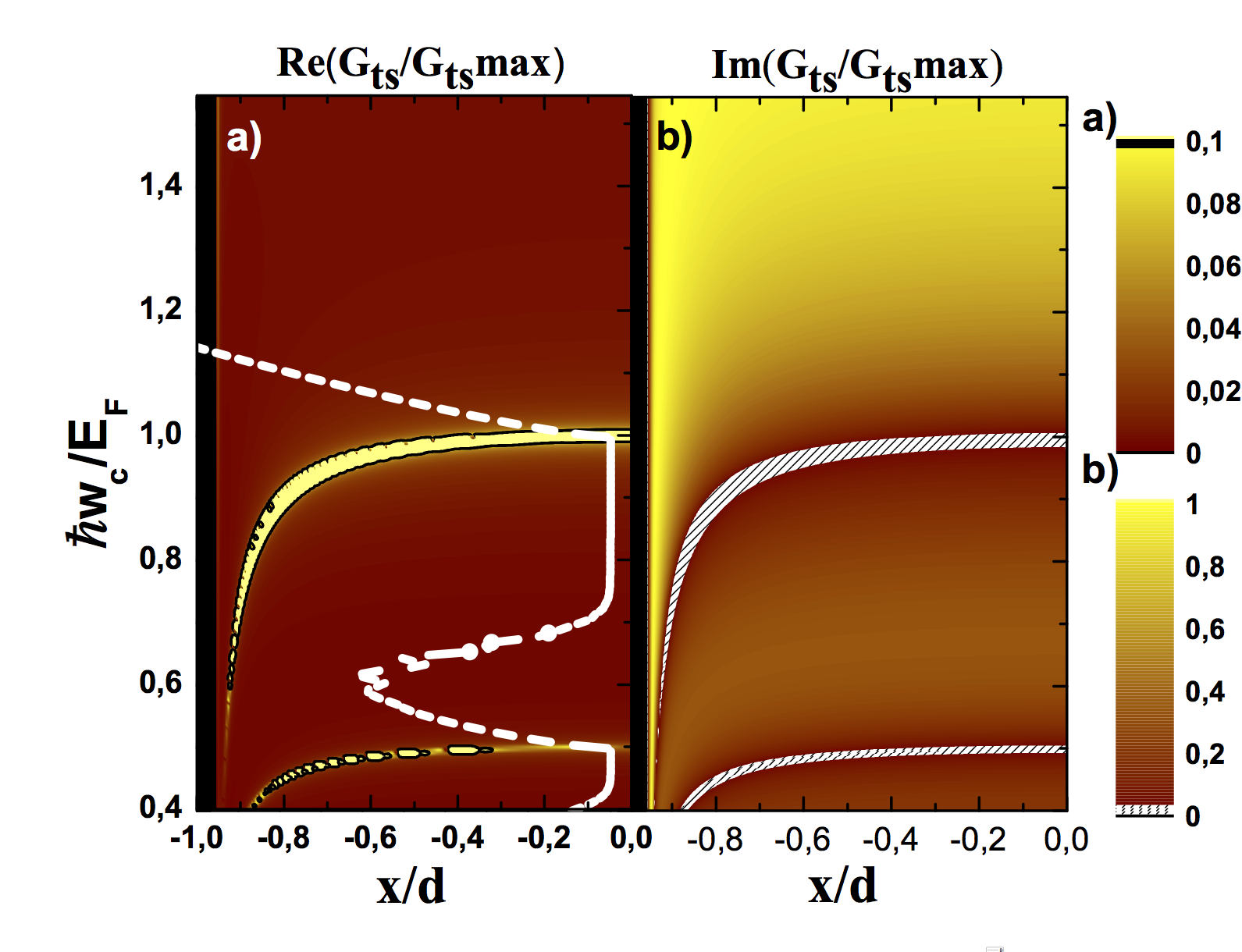}
\caption {Same gray (color) scale depicting (a) $\Re \left\lbrace G_{ts}(x)\right\rbrace $ and $\Im \left\lbrace G_{ts}(x)\right\rbrace$. The essential difference with previous two figures, here we used DOS calculated within SCBA and assuming $\gamma{=0.01}$ at default temperature $ T=2$ K. In panel (a), we also plot longitudinal resistance (broken white lines) calculated within the local equilibrium model. The interval where $R_l=0$ exactly coincides with the $B$ field intervals where a thermodynamically well developed incompressible strips/regions exist. If the system is completely compressible, $R_l$ is finite.}
\label{Fig:5}
\end{figure}

In a final numerical calculation, we investigate the effect of DOS profile on $G_{ts}(x)$, using the SCBA model. One of the most important improvement is that, we can also calculate global transport quantities, without being bothered by overlapping levels. Since, SCBA yields semi-elliptic DOS at $T=0$ and considering a high mobility sample, energy levels are well separated and a finite gap between them is obtained as a direct consequence. Even at elevated $T\lesssim10$ K, one can still observe a well defined gap~\cite{Yildiz14:014704}. Although similarities are observed between Gaussian and SCBA models presented in Fig.\ref{Fig:4} and Fig.\ref{Fig:5} apparently, as a result of the finite gaps in SCBA the narrow incompressible strips are averaged out by the thermal wavelength. 

To comprehend our investigations, we observe that using SCTFPA to obtain electron density distributions and assuming SCBA DOS model in calculating conductivities together with the Born-Oppenheimer based local Ohm's law one can almost perfectly uncover the electrostatic, thermodynamic and transport mechanisms lying beneath the experimental results reported using the DSCM technique~\cite{Suddards12:083015} among other local probe experiments~\cite{Wei98:1674,Ahlswede01:562,Dahlem10:121305}.   
\section{Conclusion}
In this report, we could describe experimental findings in a self-standing way by performing numerical calculations under well justified approaches, such as Thomas-Fermi-Poisson, Born-Oppenheimer and Self-consistent Born approximations. We observed that, assuming bare Landau quantized levels and Gaussian broadened DOS does not represent realistic samples, due to the fact that there are no ideal samples and imposed localization assumptions, respectively. It is concluded that, the direct relation between compressibility and capacitance via thermodynamical density of states exhibits itself obviously both in experiments and calculations. 

Our findings also leads us to propose further experiments to be performed using the powerful and reliable DSCM technique. As a first and relatively simple experiment would be measuring same quantities on samples with different mobilities, also at lower or higher temperatures. This may help us to understand the whether scattering broadening or temperature effects are more relevant in determining compressibility and related quantities.

\section*{Acknowledgments}
This work is financially supported by T\"UB\.{I}TAK under grant numbers 112T264 and 211T148. A.S acknowledges, Mimar Sinan Fine Arts University Administration for encouraging us to conduct a research between four different Universities.   

\section*{Referances}

\bibliography{zitate.bib}

\bibliographystyle{vancouver}
\end{document}